\documentclass[preprint,12pt]{elsarticle}

\usepackage{graphicx}
\usepackage{amsmath,amssymb}
\usepackage{xcolor}
\usepackage{float}
\usepackage{array}
\usepackage{adjustbox}
\usepackage{booktabs}
\usepackage{lineno}
\usepackage[linesnumbered,ruled,vlined]{algorithm2e}
\usepackage{subfig}

\begin{document}
\begin{frontmatter}

\title{Q-Learning–Driven Adaptive Rewiring for Cooperative Control in Heterogeneous Networks}

\author[addr1]{Yi-Ning Weng}
\author[addr2]{Hsuan-Wei Lee\corref{cor1}}
\ead{hsl324@lehigh.edu}
\cortext[cor1]{Corresponding author.}

\address[addr1]{Department of Accounting, National Taiwan University, Taiwan}
\address[addr2]{College of Health, Lehigh University, USA}

\date{}

\begin{abstract}

Cooperation emergence in multi-agent systems represents a fundamental statistical physics problem where microscopic learning rules drive macroscopic collective behavior transitions. We propose a Q-learning-based variant of adaptive rewiring that builds on mechanisms studied in the literature. This method combines temporal difference learning with network restructuring so that agents can optimize strategies and social connections based on interaction histories. Through neighbor-specific Q-learning, agents develop sophisticated partnership management strategies that enable cooperator cluster formation, creating spatial separation between cooperative and defective regions. Using power-law networks that reflect real-world heterogeneous connectivity patterns, we evaluate emergent behaviors under varying rewiring constraint levels, revealing distinct cooperation patterns across parameter space rather than sharp thermodynamic transitions. Our systematic analysis identifies three behavioral regimes: a permissive regime (low constraints) enabling rapid cooperative cluster formation, an intermediate regime with sensitive dependence on dilemma strength, and a patient regime (high constraints) where strategic accumulation gradually optimizes network structure. Comparative analysis against Bush-Mosteller stimulus-response learning demonstrates that Q-learning's temporal credit assignment capabilities produce superior cooperation outcomes, particularly under intermediate rewiring constraints where long-term relationship assessment becomes crucial. Simulation results show that while moderate constraints create transition-like zones that suppress cooperation, fully adaptive rewiring enhances cooperation levels through systematic exploration of favorable network configurations. Quantitative analysis reveals that increased rewiring frequency drives large-scale cluster formation with power-law size distributions. Our results establish a new paradigm for understanding intelligence-driven cooperation pattern formation in complex adaptive systems, revealing how machine learning serves as an alternative driving force for spontaneous organization in multi-agent networks.

\end{abstract}


\end{frontmatter}

\section{Introduction}
Ensuring cooperative control in distributed engineered systems and applications is a daunting challenge across diverse domains. In distributed resource management, cooperative agents must dynamically adapt to balance local demands and maintain global performance \cite{hao2017dynamics}; in urban traffic networks, intersections must exchange information to optimize flows \cite{salkham2008collaborative, wang2019new}; in robotic swarms, unmanned aerial vehicles or mobile robots must align actions for collective tasks under uncertainty \cite{palunko2014cooperative, li2024multi}. Apparently, in each case, the performance of the overall system, including throughput, latency, reliability, and safety, depends on the ability of autonomous agents to adapt strategies and restructure interactions in dynamic environments. Enhancing cooperation among agents is therefore essential, since insufficient coordination can lead to cascading failures, degraded performance, or even systemic collapse in critical infrastructures. This engineering perspective also mirrors a deeper scientific puzzle: how self-interested agents, across natural, social, and artificial systems, manage to sustain cooperation despite incentives to defect.

Deciphering how self-interested agents achieve cooperation remains a central open question spanning statistical physics, evolutionary biology, the social sciences, and complex systems theory \cite{perc2017statistical,sigmund1999evolutionary,friedman1991evolutionary,szabo2007evolutionary}. Despite the apparent contradiction between individual rationality and collective benefit, cooperative behavior pervades natural and artificial systems from microbial colonies \cite{frey2010evolutionary} and neural networks \cite{bichler2021learning} to economic markets \cite{li2023evolutionary} and distributed computing architectures \cite{du2022sdn}. This ubiquity suggests that cooperation emergence follows universal physical principles that transcend specific biological or technological implementations. Understanding these fundamental mechanisms has profound implications for designing beneficial artificial societies and explaining the ubiquity of cooperation in nature.

The Prisoner's Dilemma (PD) has served as the canonical framework for investigating cooperation emergence, where individual rationality paradoxically leads to collectively suboptimal outcomes. Traditional studies of PD have primarily considered static networks or well-mixed populations, where agents interact with fixed partners over repeated rounds or are distributed in homogeneous topologies \cite{nowak1992evolutionary, nowak2004evolutionary, lieberman2005evolutionary, perc2013evolutionary, szolnoki2018evolutionary, lee2022mercenary, lee2024supporting}. These classical approaches revealed fundamental cooperation mechanisms including spatial reciprocity, group selection, and reputation effects. However, real-world systems exhibit inherently dynamic topologies with continuously evolving connections and heterogeneous agent behaviors. This dynamic nature suggests that network plasticity represents a fundamental driver of cooperation emergence rather than a secondary supportive factor.

Recognition of this limitation has sparked extensive research into adaptive networks that grant agents the freedom to break unproductive ties and forge new ones. In recent years, researchers have extended the scope of PD studies to dynamic networks, allowing agents to adjust their connections over time through tie breaking and edge formation \cite{fu2009partner, lee2018evolutionary, su2023strategy, traulsen2023future}. Such network rewiring mechanisms offer agents the ability to sever unproductive links and form new ties, mimicking the adaptive nature of social relationships \cite{gross2008adaptive, pinheiro2016linking, malik2016transitivity, lee2019social}. Theoretical and empirical studies have demonstrated that even simple rewiring rules, such as preferentially connecting to cooperators or severing ties with defectors, can dramatically improve population-level cooperation \cite{zhang2012different, melamed2016strong, pan2018evolution, melamed2018cooperation}. These findings established network adaptability as a powerful mechanism for escaping exploitation traps and forming stable cooperative clusters.

Despite these advances, most rewiring mechanisms rely on predetermined heuristics or static strategies that lack the flexibility to adapt to diverse local environments through experience. Recent advances have shown that incorporating reinforcement learning into network rewiring decisions can significantly enhance cooperation levels, with adaptive rewiring mechanisms guided by Bush-Mosteller learning \cite{bush1953stochastic, macy2002learning, masuda2011numerical} creating more stable cooperative structures compared to static strategies \cite{han2024analysis, lee2025enhancing}. However, these approaches typically employ simple stimulus-response learning without the sophisticated temporal credit assignment capabilities of modern reinforcement learning algorithms.

Recent advances in machine learning now allow these adaptive rules to be learned rather than hand-coded, linking statistical-physics models to modern artificial intelligence \cite{bloembergen2015evolutionary, han2022emergent, wang2022cooperative}. Reinforcement learning (RL) has emerged as particularly powerful for designing adaptive agents capable of learning optimal strategies through environmental feedback \cite{jia2021local, pu2022attention, geng2022reinforcement, song2022reinforcement, du2024evolution, mintz2025evolutionary}. Q-learning is particularly attractive because it is model-free, converges under mild conditions, and scales to the large heterogeneous graphs typical of social systems \cite{shi2022analysis, dolgopolov2024reinforcement,yang2024interaction, zheng2024evolution, lin2025coevolution, zhu2025q}. The temporal difference learning mechanism in Q-learning enables sophisticated evaluation of long-term relationship value, making it ideally suited for partnership management in dynamic social networks.

Complementary work on Interactive Diversity (ID) shows that agents gain when they tailor their actions to each neighbor rather than deploying a single global strategy \cite{su2016interactive, su2017evolutionary, su2018evolution, su2019evolutionary, si2025evolution}. Edge-based analytical frameworks reveal that such personalized interaction strategies are crucial for maintaining extensive reciprocal relationships, ultimately generating resilient cooperative clusters that persist across diverse parameter regimes. This Interactive Diversity framework has demonstrated superior performance in sustaining cooperation compared to traditional uniform-strategy approaches, particularly under moderate temptation conditions \cite{lee2025granular}. However, the potential for combining such behavioral sophistication with intelligent network adaptation through reinforcement learning remains largely unexplored.

Bringing value-based learning together with neighbor-specific decision making therefore offers an unexplored route to agents that optimize both behavior and topology in tandem. While extensive research has focused on learning action strategies (deciding when to cooperate or defect), the potential for reinforcement learning to guide intelligent network rewiring decisions remains largely unexplored. To address this limitation, we propose a novel framework that integrates Q-learning with adaptive network rewiring in dynamic Prisoner's Dilemma games. Our approach introduces two key innovations that bridge microscopic learning dynamics with macroscopic cooperation emergence. First, building upon the Interactive Diversity framework \cite{lee2025granular}, we introduce agents that independently determine distinct strategies for different neighbors rather than applying uniform policies. Second, we implement dual-layer Q-learning where agents simultaneously learn optimal action policies and rewiring decisions, creating an integrated adaptation mechanism that operates across both behavioral and structural dimensions.

Our framework enables agents to learn not only how to act but also with whom to interact, guided by Q-learning algorithms that evaluate long-term relationship value through temporal difference learning. Each time the rewiring constraint (RC) is satisfied, agents use temporal difference learning to assess whether maintaining connections with specific neighbors remains beneficial, potentially rewiring to partners with compatible action preferences. This creates a coevolutionary dynamic where behavioral learning and network structure mutually influence each other, enabling systematic exploration of both strategy space and network configuration space. By scanning the rewiring constraint $RC$ we reveal three dynamical regimes—permissive, critical, and patient—that mirror an order-disorder transition controlled by $RC$ rather than temperature. Our results demonstrate that Q-learning-based rewiring substantially enhances both global cooperation levels and structural organization compared to heuristic and Bush-Mosteller strategies, establishing new paradigms for understanding how machine learning drives collective behavior in adaptive social networks.

\section{Methods}

Our model integrates Q-learning with adaptive network rewiring in the Prisoner's Dilemma Game (PDG), building upon established frameworks for reinforcement learning in evolutionary games \cite{lee2025enhancing, geng2022reinforcement, lee2025granular}. This approach enables agents to simultaneously learn optimal strategies and adaptively restructure their social connections based on interaction outcomes, bridging microscopic learning dynamics with macroscopic collective behavior emergence. The framework represents a novel synthesis of temporal difference learning and network evolution, where agents optimize both behavioral policies and social partnerships through experience-based adaptation.

\subsection{Game Structure and Network Topology}

We employ a parameterized Prisoner's Dilemma payoff matrix that systematically captures varying dilemma intensities across different classes of social conflicts. Following established frameworks that integrate dilemma strength theory with evolutionary game dynamics \cite{geng2022reinforcement, lee2025granular, wang2013insight}, our payoff structure is expressed as:
\begin{equation}\label{eq:matrix}
\begin{pmatrix}
R & S \\
T & P
\end{pmatrix}
=
\begin{pmatrix}
1 & -D_r \\
1 + D_g & 0
\end{pmatrix}
\end{equation}
where $D_g = T - R \, (0 \leq D_g \leq 1)$ quantifies the temptation advantage for unilateral defection, representing the chicken-type dilemma component, and $D_r = P - S \, (0 \leq D_r \leq 1)$ measures the punishment severity for being exploited, corresponding to the stag-hunt-type dilemma element \cite{wang2013insight}. We adopt the dilemma strength, quantified by $D_g$ and $D_r$, to characterize the severity of social dilemmas, as originally formulated in the seminal works on weakly dominant strategies in symmetric games \cite{tanimoto2007relationship}, universal scaling for dilemma strength \cite{wang2015universal}, and the phase-plane scaling approach for game-class transitions \cite{ito2018scaling}. This parameterization with fixed $R = 1$ and $P = 0$ establishes a normalized monetary scale while enabling systematic exploration of cooperation dynamics across the complete space of two-player social dilemmas through variation of $(D_r, D_g) \in [0, 1]^2$. In this framework, $D_r$ primarily controls the risk of exploitation while $D_g$ modulates the reward for defection \cite{wang2013insight}. The parameter space $(D_r, D_g)$ effectively maps different classes of social dilemmas: when $D_r > D_g$, the system exhibits stag-hunt-like characteristics emphasizing coordination benefits, while $D_g > D_r$ creates chicken-game dynamics where temptation dominates punishment avoidance. Consistent with prior findings that cooperation exhibits greater sensitivity to $D_r$ than to $D_g$ \cite{geng2022reinforcement}, we focus our analysis on the diagonal constraint $D_g = D_r$, systematically varying $D_r \in [0, 0.3]$. This dimensional reduction from the full $(D_r, D_g)$ parameter space is essential for computational tractability in our extensive simulations involving up to $10^7$ time steps on power-law networks, while comprehensive exploration of the complete two-dimensional parameter plane remains a valuable direction for future investigations.

We employ power-law networks with degree distribution $P(k) \propto k^{-\gamma}$ where $\gamma = 3$, accurately reflecting the heterogeneous connectivity patterns observed in real-world social and biological systems \cite{barabasi2013network, newman2018networks}. Each network consists of $N$ nodes with an average degree of $\langle k \rangle = 4$, chosen to balance computational tractability with sufficient connectivity for meaningful social interactions while avoiding percolation threshold effects that could artificially enhance cooperation. Networks are generated using the configuration model, which preserves the desired degree sequence while randomizing connections, ensuring that observed cooperation patterns emerge from learning dynamics rather than structural biases. This scale-free topology choice reflects empirical evidence that social networks exhibit power-law degree distributions, where highly connected hubs coexist with sparsely connected nodes, fundamentally altering cooperation dynamics compared to homogeneous structures such as regular lattices or random graphs.

Agents are initialized with random initial behavioral tendencies toward cooperation or defection with equal probability ($p_{\text{init}} = 0.5$), before developing neighbor-specific strategies through learning. We also ensure that each agent has at least two neighbors to maintain fairness in the rewiring opportunities. The initial rewiring decision is set to ``Not Rewire''. This initialization protocol ensures that the system begins in a completely mixed state without any initial bias toward cooperation or defection, allowing us to observe genuine emergence of cooperative structures through learning processes rather than initial condition effects.

\subsection{Neighbor-Specific Q-Learning Framework}

Following the Interactive Diversity (ID) framework \cite{lee2025granular}, our agents implement neighbor-specific decision-making rather than applying uniform strategies across all connections. This granular approach recognizes that real-world social interactions often involve tailored responses to different relationship partners, enabling more sophisticated behavioral adaptation than monolithic strategies. The ID framework represents a significant departure from traditional evolutionary game theory, where agents typically employ identical strategies against all opponents, by acknowledging that intelligent agents can maintain distinct behavioral policies for different social relationships simultaneously.

The temporal dynamics follow two-timescale separation: fast action adaptation ($\tau_{\text{action}} = 1$) with synchronous Q-value updates, and slower structural evolution ($\tau_{\text{rewire}} = RC$) with rewiring decisions updated every $RC$ rounds based on accumulated rewards. Each agent with $n$ neighbors maintains $n$ parallel Q-learning processes, one for each neighbor relationship, enabling different actions toward different neighbors within the same time step. Payoffs are determined by Equation (\ref{eq:matrix}) entries, with each agent calculating payoffs separately for each neighbor. This multi-timescale architecture reflects the natural hierarchy observed in social systems, where behavioral adjustments occur more rapidly than structural changes to social networks, creating a realistic framework for studying coevolutionary dynamics.

\textbf{Action Selection:} We employ Q-learning as the core reinforcement learning mechanism for granular action selection, where each agent independently evaluates the state of each neighbor and may adopt different actions for different neighbors within the same time step. For an agent with $n$ neighbors choosing action $a_i$ in state $s_i$, Q-values are updated as:
\begin{equation}\label{eq:q-ID}
AQ_{t+1}(s_i, a_i) = AQ_t(s_i, a_i) + \alpha \left[ r_t(s_i, a_i) + \gamma \max_{a' \in A} AQ_t(s_i', a') - AQ_t(s_i, a_i) \right]
\end{equation}
where $\alpha = 0.1$ is the learning rate, $\gamma = 0.9$ is the discount factor, and $r_t$ represents immediate rewards from payoff matrix entries. The state space $\mathcal{S}_{\text{AQ}}^{(i)} = \{0, 1, 2\}$ encodes the number of cooperating agents in each interaction pair: $s_i = 0$ indicates both agents defected, $s_i = 1$ represents mixed behavior (one cooperates, one defects), and $s_i = 2$ signifies mutual cooperation. This minimal state representation captures the essential cooperative context while maintaining computational tractability, effectively encoding the local interaction history that determines optimal action selection. The binary action set $A = \{0, 1\}$ represents cooperation and defection.

We set Q-learning parameters based on established best practices in reinforcement learning literature and sensitivity analysis: $\alpha = 0.1$ provides moderate learning that balances adaptation speed with stability, $\gamma = 0.9$ emphasizes future rewards while maintaining responsiveness to immediate outcomes, and the temporal difference mechanism enables sophisticated evaluation of long-term relationship benefits beyond immediate payoff maximization \cite{geng2022reinforcement, lee2025granular}. Both action selection and rewiring decisions follow $\varepsilon$-greedy policies with $\varepsilon = 0.02$, ensuring minimal exploration while maintaining exploitation of learned strategies. This stochastic policy ensures ergodicity while creating a balance between deterministic optimization and adaptive exploration, preventing the system from becoming trapped in suboptimal behavioral patterns.

\textbf{Adaptive Rewiring:} Network rewiring represents a critical mechanism for escaping exploitation and forming beneficial partnerships, effectively allowing the system to explore different regions of configuration space in search of cooperative equilibria. Agents eligible for rewiring (those who cooperated while neighbors defected) evaluate connection-breaking decisions using Q-learning and may reconnect to agents with similar action preferences, implementing homophily-based assortative mixing. This conditional rewiring mechanism prevents purely random network changes while allowing strategic relationship adjustment based on interaction outcomes, creating directed evolution toward more cooperative network structures.

The rewiring constraint parameter $RC$ controls temporal scale: larger values promote strategic patience by requiring agents to accumulate more interaction history before making structural changes, while smaller values enable rapid network adaptation. This creates a tunable separation of timescales between action learning ($\tau_{\text{action}} = 1$) and structural evolution ($\tau_{\text{rewire}} = RC$), enabling systematic exploration of how different temporal hierarchies affect cooperation emergence.

For rewiring decision $rd_i$ in state $s_i$:
\begin{equation}\label{eq:q-Rewire}
RQ_{t+1}(s_i, rd_i) = RQ_t(s_i, rd_i) + \alpha \left[ pr_t(s_i, rd_i) + \gamma \max_{rd' \in RD} RQ_t(s_i', rd') - RQ_t(s_i, rd_i) \right]
\end{equation}
where $pr_t$ represents cumulative rewards over $RC$ rounds, capturing long-term relationship value. Unlike the immediate rewards used for action learning, these cumulative rewards enable agents to assess relationship quality over extended time horizons, distinguishing between partners who provide consistent benefits versus those who exploit cooperative gestures. The state space $\mathcal{S}_{\text{RQ}}^{(i)} = \{0, 1, 2\}$ mirrors action learning, while the decision set $RD = \{0, 1\}$ represents ``Rewire'' and ``Not Rewire'' choices.

For homophily-based reconnection, if an agent has more than 50\% cooperative actions, it is classified as a cooperator and attempts to establish new connections with other cooperators; conversely, agents with predominantly defective actions seek connections with other defectors. This assortative mixing reflects empirical tendencies in social networks and can facilitate cooperative cluster formation and stabilization.

\subsection{Bush-Mosteller Baseline Model}

To compare the influence of reinforcement learning on rewiring decisions, we introduce the Bush-Mosteller (BM) model \cite{lee2025enhancing} as a baseline that represents stimulus-response learning without explicit value function approximation. The BM model implements a form of associative learning where agents adjust their rewiring propensities based on immediate satisfaction or dissatisfaction with recent outcomes, without the temporal credit assignment capabilities of Q-learning.

An agent's probability of breaking a connection, $p_t$, depends on its stimulus $s_t = \tanh[\beta(r_t - A)]$, where $A$ is the constant aspiration level, $r_t$ is the periodic payoff from a specific neighbor, and $\beta$ controls the sensitivity. The parameter $\beta$ modulates how strongly agents react to payoff deviations from their aspirations: higher values create more decisive responses to satisfaction or dissatisfaction, while lower values promote gradual behavioral adjustments, creating a tunable response function that ranges from linear ($\beta \to 0$) to step-like ($\beta \to \infty$) behavior. Each time the network accumulates $RC$ rounds, agents update their breaking probability based on:

\begin{equation}\label{eq:BM}
p_t = 
\begin{cases}
p_{t-1} + (1 - p_{t-1}) s_{t-1} & \text{if } rd_{t-1} = C,\ s_{t-1} \geq 0 \\
p_{t-1} + p_{t-1} s_{t-1} & \text{if } rd_{t-1} = C,\ s_{t-1} < 0 \\
p_{t-1} - p_{t-1} s_{t-1} & \text{if } rd_{t-1} = D,\ s_{t-1} \geq 0 \\
p_{t-1} - (1 - p_{t-1}) s_{t-1} & \text{if } rd_{t-1} = D,\ s_{t-1} < 0
\end{cases}
\end{equation}

where $rd_{t-1}$ denotes the agent's previous rewiring decision. The probability $p_t$ is reinforced if the stimulus $s_{t-1}$ is positive and diminished if $s_{t-1}$ is negative, with deviation probability $\varepsilon = 0.2$ for occasional random decisions. This implements the Law of Effect \cite{postman1947history} through probabilistic reinforcement, where satisfying outcomes increase the likelihood of repeating successful behaviors while disappointing results reduce their probability, providing a psychologically motivated baseline for comparison with Q-learning approaches.

We set BM model parameters based on behavioral learning principles and empirical considerations. An inverse temperature parameter \(\beta = 2\) balances responsiveness and stochasticity, ensuring that agents exhibit probabilistic but not overly deterministic behavior in response to payoff differences. The aspiration update factor \(A = 1.1\) allows agents to rapidly adjust their aspiration levels toward recent outcomes, maintaining sensitivity to environmental changes while retaining a minimal degree of temporal smoothing.

\subsection{Simulation Protocol and Statistical Analysis}

We conduct simulations on power-law networks with $N \in \{10^4, 2 \times 10^4, 5 \times 10^4, 10^5\}$ nodes, ensuring statistical robustness while avoiding finite-size effects that could bias cooperation dynamics in small populations. Finite-size scaling analysis confirms that cooperation levels converge to well-defined thermodynamic limits for $N \geq 20{,}000$. Each simulation typically runs for $10^5$ time steps, extended to $10^7$ time steps for special cases to ensure equilibrium behavior. We perform 30 independent trials for each parameter combination using different random seeds to account for stochastic variability. These extended simulations are particularly important for parameter regimes near critical transitions, where relaxation times can become extremely long and systems may exhibit slow approach to equilibrium.

We measure cooperation by the fraction of cooperative actions executed, providing an action-based order parameter that directly reflects behavioral dynamics rather than agent classifications, which can be ambiguous in systems with mixed strategies. To capture long-term behavior, we compute the average fraction of cooperative actions over the final 5{,}000 time steps of each run, after transient dynamics have subsided. This measurement window is chosen to ensure system equilibration while providing sufficient statistical sampling of steady-state behavior, with analysis confirming that cooperation levels reach stable values within the first $5 \times 10^4$ time steps.

\begin{equation}\label{eq:ave}
\text{Cooperation Level} = \frac{\text{Total number of cooperative actions in the final 5{,}000 steps}}{4 \times N \times 5{,}000}
\end{equation}

This cooperation level serves as our primary order parameter, analogous to magnetization in spin systems, enabling quantitative analysis of cooperation patterns and transition-like behaviors. The time-averaged value from Equation~(\ref{eq:ave}), together with the contemporaneous mean payoff per agent, serves as our quantitative measure of collective welfare and system performance.

\section{Results}

We investigate how Q-learning-driven adaptive rewiring transforms cooperation dynamics in multi-agent systems, revealing emergent changes in cooperation regimes that bridge microscopic learning rules with macroscopic collective behavior. Our primary goal is to evaluate whether agents endowed with the ability to adapt both their strategies and their connections, through independent Q-learning updates, can foster stable and widespread cooperation on evolving power-law networks. To this end, we systematically varied the level of rewiring constraint (RC) and compared the outcomes against several baseline models, including random rewiring mechanism and other reinforcement learning rewiring strategies such as the BM model. Through systematic parameter sweeps and statistical analysis, we uncover universal patterns in cooperation emergence under adaptive rewiring and show how incorporating Q-learning modifies these patterns, particularly in intermediate constraint regimes where long-term partner evaluation becomes advantageous. Our analysis reveals three distinct dynamical regimes: a permissive regime (low $RC$) enabling rapid cooperative cluster formation, an intermediate regime (intermediate $RC$) with sensitive dependence on dilemma strength, and a patient regime (high $RC$) where strategic accumulation gradually optimizes network structure.

\subsection{Cooperation Landscape: Regime Changes in Parameter Space}

\begin{figure}[H]
   \centering
   \includegraphics[width=0.9\textwidth]{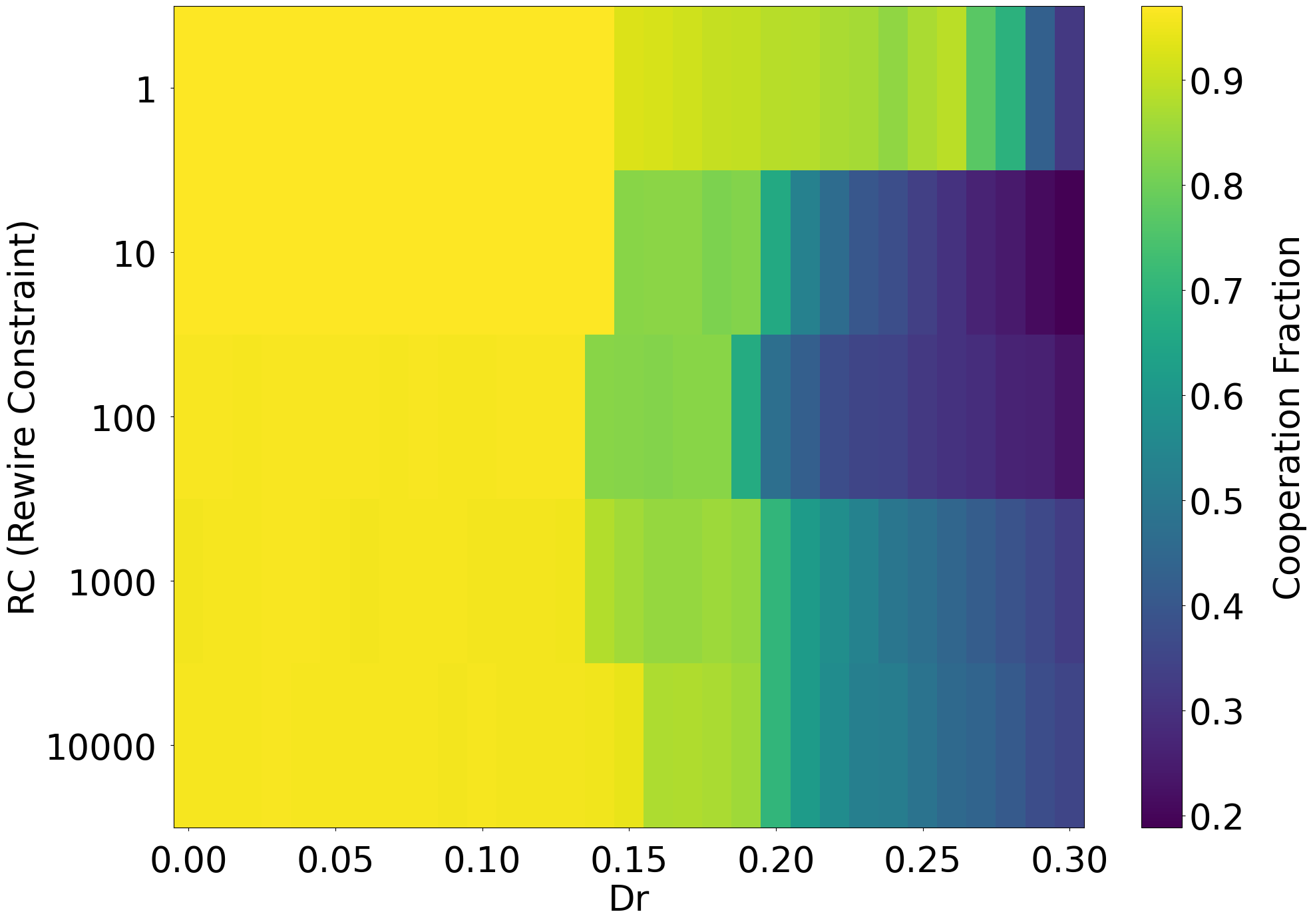}
   \caption{Steady-state fraction of cooperators in the $(RC, D_r)$ parameter space. Heatmap of average cooperation levels in the stationary states across different values of rewiring constraint ($RC$) and dilemma strength ($D_r$). Simulations were conducted on a power-law network of size $N = 50{,}000$ for $10^7$ time steps, with each data point averaged over 30 independent runs. The diagram reveals distinct regimes: a permissive regime (high cooperation for low $RC$), an intermediate regime (sensitive to $D_r$), and a patient regime (high $RC$ with conditional cooperation). Boundary regions mark where cooperation undergoes pronounced changes, reminiscent of order–disorder patterns in statistical physics but without evidence of a sharp phase transition. The figure highlights that high cooperation levels are sustained under low $D_r$ and low $RC$, while intermediate constraint levels exhibit significant drops in cooperation under harsher dilemma conditions.}
   \label{fig:heatmap}
\end{figure}

To understand the relationship between cooperation level and rewiring constraints, we conducted large-scale simulations across different values of $RC$ (rewiring constraint) and $D_r$ (sucker's payoff loss), with $N = 50{,}000$ agents and each configuration simulated for $10^7$ time steps. Figure~\ref{fig:heatmap} presents a parameter-space map summarizing the average cooperation level in the stationary states for each $(RC, D_r)$ pair, computed over 30 independent runs to ensure robustness.

The results reveal a rich structure of cooperation regimes with non-monotonic shifts between rewiring flexibility and cooperation dynamics. When $RC = 1$, indicating full rewiring freedom, the system achieves high cooperation rates exceeding $90\%$ for most values of $D_r$ up to approximately $0.18$. Beyond this threshold, cooperation gradually declines but still remains higher than in more constrained rewiring conditions, underscoring the effectiveness of adaptive network restructuring in supporting prosocial behavior. This regime represents a predominantly cooperative state where rapid structural adaptation enables escape from local exploitation traps. As the constraint level increases (e.g., $RC = 100$), we observe a sharp decline in cooperation, particularly when $D_r > 0.15$, suggesting that limiting rewiring opportunities disrupts agents' ability to escape exploitative ties. This defines a boundary region separating cooperative and mixed regimes in parameter space, where network heterogeneity creates degree-dependent cooperation patterns that vary systematically across rewiring regimes (Figure~\ref{fig:degree_dependent_cooperation}). Interestingly, for extremely high constraints ($RC = 10{,}000$), cooperation partially rebounds at low $D_r$ but remains suppressed in more hostile environments. This U-shaped pattern indicates a re-entrant behavior where both overly flexible and highly constrained systems can sustain cooperation through different mechanisms: rapid adaptation versus long-term strategic patience.

The parameter-space map reveals three distinct regimes with characteristic cooperation dynamics: (i) a permissive regime for low $RC$ values where high cooperation emerges across all $D_r$ conditions through rapid structural adaptation, (ii) an intermediate regime where cooperation depends sensitively on dilemma strength and exhibits pronounced changes, and (iii) a patient regime for high $RC$ values where cooperation recovers only under mild dilemma conditions through long-term strategic accumulation. These findings illustrate the crucial role of network plasticity in modulating cooperative behavior and highlight the need for balanced rewiring mechanisms in reinforcement learning-driven multi-agent systems.

\subsection{Temporal Dynamics and Relaxation Processes}

\begin{figure}[H]
   \centering
   \includegraphics[width=0.9\textwidth]{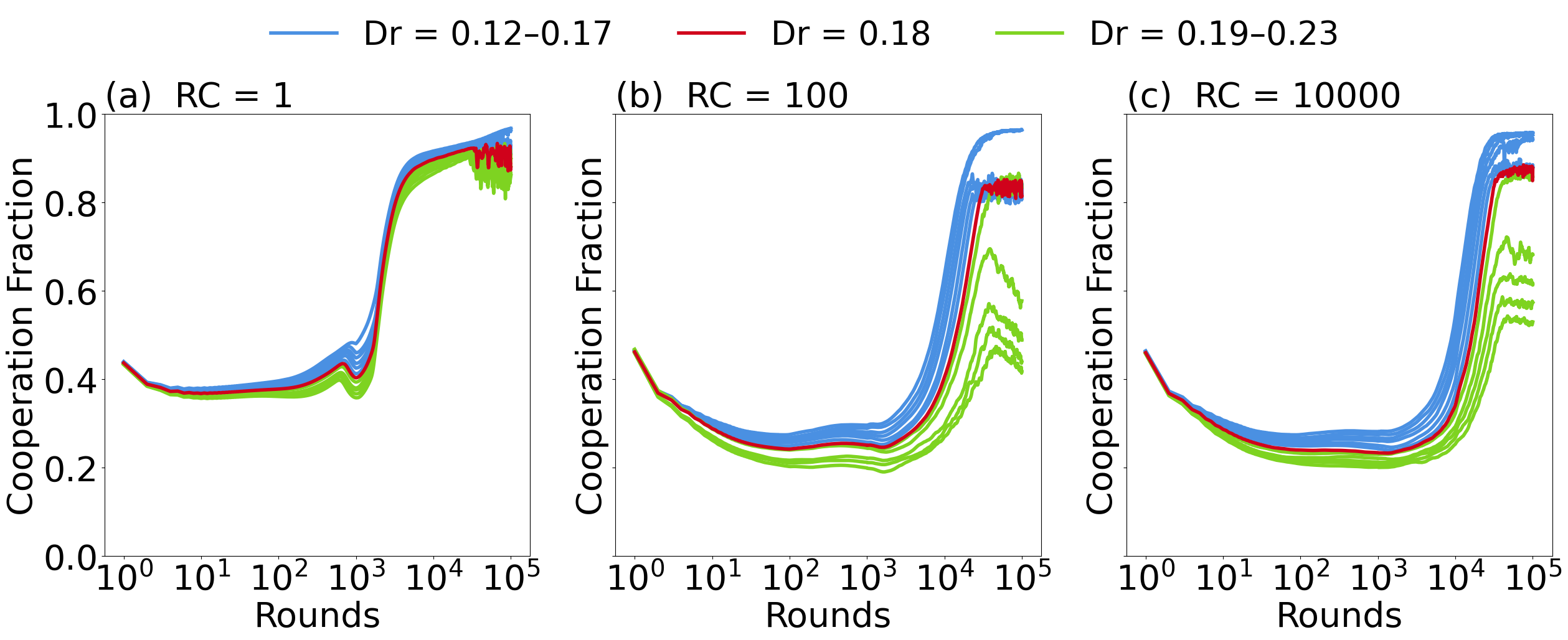}
   \caption{Relaxation dynamics and approach to equilibrium under different constraint regimes. Time evolution of cooperation levels under different rewiring constraints. Each panel shows the average cooperation rate across $10^5$ simulation rounds under (a) $RC = 1$, (b) $RC = 100$, and (c) $RC = 10{,}000$. Each subplot corresponds to a distinct RC value and includes multiple curves spanning different $D_r$ values. Specifically, blue lines correspond to $D_r = 0.12$--$0.17$, red lines to $D_r = 0.18$, and green lines to $D_r = 0.19$--$0.23$. Distinct relaxation timescales emerge: rapid convergence for the permissive regime ($\tau \sim 10^4$), non-monotonic dynamics for the critical regime ($\tau \sim 10^4$--$10^5$), and slow approach for the patient regime ($\tau > 10^5$). While most configurations converge within $10^5$ rounds, certain parameter sets (e.g., $D_r = 0.12$--$0.14$ for $RC = 1$ and $D_r = 0.20$--$0.23$ for $RC = 100$) have not yet reached equilibrium. For these cases, we provide extended results up to $10^7$ rounds in Appendix~A to confirm convergence.}
   \label{fig:coop_time}
\end{figure}

To further examine the temporal dynamics of cooperation, Figure~\ref{fig:coop_time} illustrates the cooperation rate over time under three representative levels of rewiring constraint: $RC = 1$, $RC = 100$, and $RC = 10{,}000$. Each subplot corresponds to a distinct RC value and includes multiple curves spanning different $D_r$ values. Specifically, blue lines correspond to $D_r = 0.12$--$0.17$, red lines to $D_r = 0.18$, and green lines to $D_r = 0.19$--$0.23$. These trajectories reveal the underlying relaxation processes and characteristic timescales governing approach to equilibrium.

The results reveal distinct trajectories toward equilibrium under different constraint levels. In the permissive regime ($RC = 1$), where agents can rewire freely, cooperation levels steadily increase and converge rapidly to near 1.0, even when $D_r$ is relatively high. The rapid convergence ($\tau_{\text{relax}} \sim 10^4$ time steps) indicates efficient exploration of configuration space enabled by frequent rewiring. In contrast, the critical regime ($RC = 100$) exhibits a significant decline in cooperation during early stages, with recovery occurring only after roughly $10^4$ rounds. This non-monotonic relaxation suggests competing dynamics between exploitation prevention and cluster formation, with intermediate timescales creating transient traps. This lag is especially pronounced for $D_r = 0.20$, $0.21$, $0.22$, and $0.23$, where the system has not yet reached equilibrium by $10^5$ rounds. A similar delayed stabilization is observed for $RC = 1$ with $D_r = 0.12$, $0.13$, and $0.14$. In the patient regime ($RC = 10{,}000$), the system remains at a low-cooperation state for most parameter settings, with only modest recovery occurring late in the simulation through gradual strategic optimization. Analysis of relaxation times suggests subdiffusive exploration of network configuration space as rewiring becomes constrained. 

To clarify the role of payoff cost, we categorize $D_r$ values into three colored groups: blue ($D_r = 0.12$--$0.17$), red ($D_r = 0.18$), and green ($D_r = 0.19$--$0.23$). The blue group consistently leads to high final cooperation levels across all $RC$ values, suggesting that these conditions support robust prosocial behavior. The red group lies near a critical threshold: cooperation is sustained under low constraint but begins to degrade as rewiring becomes limited. In contrast, the green group fails to recover cooperation under medium and high constraint levels, indicating a collapse in cooperative dynamics when the cost of defection is too high. These patterns highlight how payoff structure and rewiring flexibility jointly determine the emergence and stability of cooperation.

\subsection{Degree-Dependent Cooperation under Rewiring Constraints}

\begin{figure}[H]
   \centering
   \includegraphics[width=0.9\textwidth]{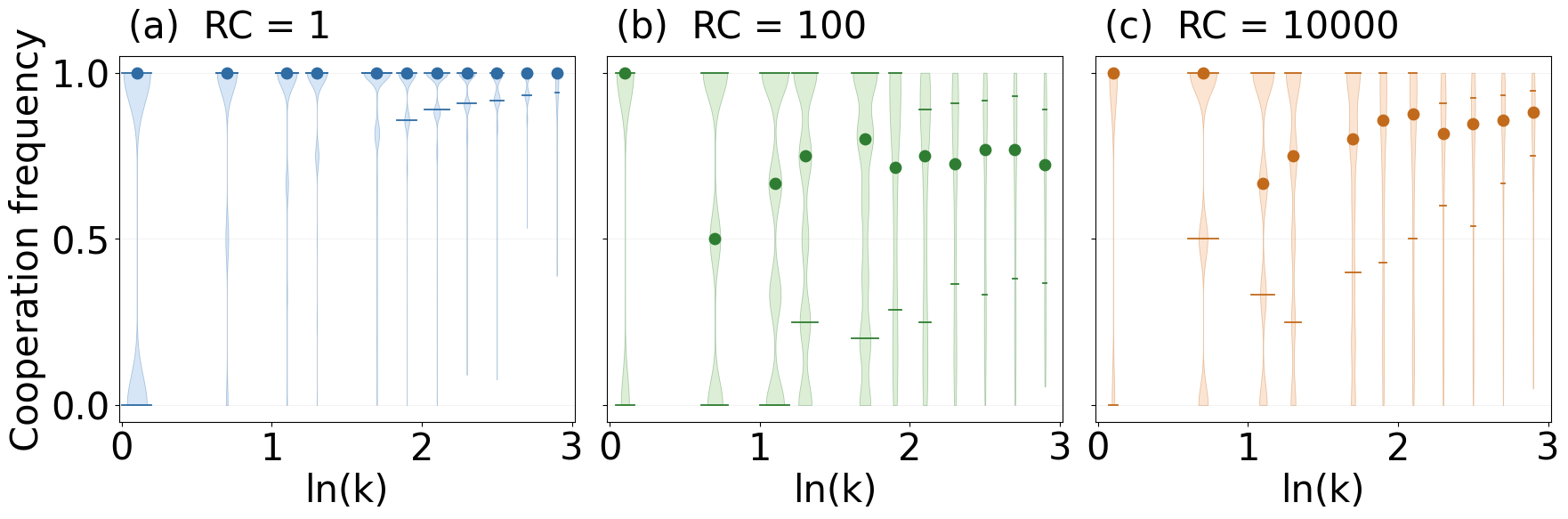}
   \caption{Degree-dependent cooperation frequency under varying rewiring constraints. These violin plots show the distribution of cooperation frequencies across agents with different degrees (log-scaled) for (a) $RC = 1$, (b) $RC = 100$, and (c) $RC = 10,000$. Dots indicate median values and horizontal bars represent interquartile ranges. Results demonstrate bimodal stratification with cooperative hubs under low RC, frustrated mixing under intermediate RC, and partial recovery of cooperation at high-degree nodes under strong constraints. These findings reveal how network heterogeneity and rewiring flexibility jointly shape cooperation outcomes, with implications for the design of resilient adaptive multi-agent systems.}
   \label{fig:degree_dependent_cooperation}
\end{figure}

To further investigate how network heterogeneity interacts with adaptive rewiring, Figure~\ref{fig:degree_dependent_cooperation} illustrates the relationship between node degree  and cooperation frequency across different constraint levels. When rewiring is fully permissive ($RC = 1$), cooperation emerges robustly across low-, medium-, and high-degree nodes, with hubs converging to full cooperation and acting as anchors that stabilize the network. This indicates that frequent rewiring not only consolidates cooperative clusters but also 
prevents peripheral agents from being trapped in exploitative ties. 

Under intermediate constraints ($RC = 100$), cooperation becomes highly dispersed with sharp degree-dependent variation. Low-degree nodes in particular exhibit significant declines in cooperation frequency, reflecting frustrated local dynamics where agents outside large clusters struggle to maintain cooperation. This highlights the vulnerability of sparsely connected agents when rewiring opportunities are limited, as they lack the structural flexibility to escape defection traps. 

In contrast, when rewiring opportunities are extremely limited ($RC = 10,000$), cooperation partially recovers among high-degree nodes. This aligns with our earlier findings, indicating that hubs can gradually consolidate cooperative ties through patient accumulation of beneficial relationships. The overall pattern suggests a non-monotonic effect of rewiring constraints: while both very frequent and very rare rewiring stabilize cooperation, intermediate levels create a critical regime where cooperation becomes fragile and strongly degree-dependent. 

\subsection{Microscopic Drift and Hazard Ratios across Node Degrees}

\begin{figure}[H]
   \centering
   \includegraphics[width=0.6\textwidth]{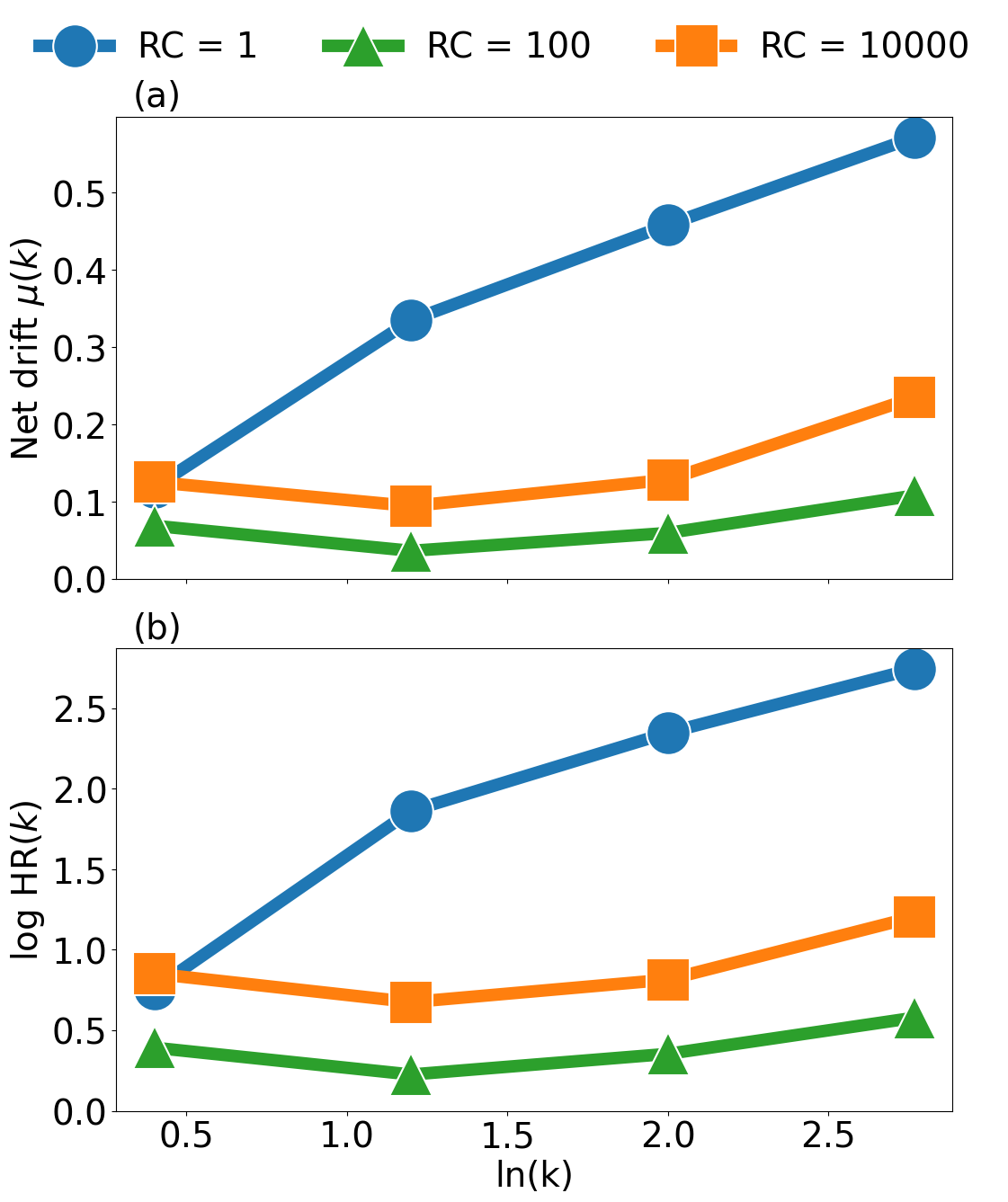}
   \caption{Microscopic drift and hazard ratios across node degrees under varying rewiring constraints. (a) Net drift $\mu(k)$ of cooperative state transitions as a function of degree $\ln(k)$. Positive values indicate a bias toward cooperation. (b) Log hazard ratio $HR(k)$ of defection risk relative to cooperation persistence. Results are shown for RC = 1 (blue circles), RC = 100 (green triangles), and RC = 10,000 (orange squares). High-degree nodes exhibit strong cooperative drift and reduced defection risk under permissive rewiring, while intermediate constraints suppress both effects. These findings confirm that network hubs play a central role in stabilizing cooperation and suggest design strategies for robust adaptive multi-agent systems.}
   \label{fig:drifthr}
\end{figure}

To quantify the microscopic mechanisms underlying degree-dependent cooperation, 
Figure~\ref{fig:drifthr} presents the net drift $\mu(k)$ and the hazard ratio $HR(k)$ of cooperative behavior as functions of node degree under different rewiring constraints. 
The drift metric captures the directional bias in cooperative state transitions, while 
the hazard ratio measures the relative risk of defection versus cooperation persistence. 

Formally, the net drift is defined as
\begin{equation}
\mu(k) = h_{D \to C}(k) - h_{C \to D}(k),
\end{equation}
where $h_{D \to C}(k)$ denotes the hazard rate of transitions from defection to cooperation 
and $h_{C \to D}(k)$ denotes the hazard rate of transitions from cooperation to defection. 

The hazard rate of defection relative to cooperation is given by
\begin{equation}
HR(k) = \frac{h_{D \to C}(k)}{h_{C \to D}(k)},
\end{equation}
and we additionally consider its logarithmic form $\log HR(k)$ to amplify subtle differences in transition dynamics. 

Here the elementary hazard terms are computed as
\begin{equation}
h_{D \to C}(k) = \frac{N_{D \to C}(k)}{T_D(k)}, \qquad
h_{C \to D}(k) = \frac{N_{C \to D}(k)}{T_C(k)},
\end{equation}
where $N_{D \to C}(k)$ is the number of observed transitions from defection to cooperation for nodes of degree $k$, 
$T_D(k)$ is the total number of time steps in which nodes of degree $k$ are in the defective state, 
and the definitions for $h_{C \to D}(k)$ follow analogously.

Panel (a) shows that under permissive rewiring ($RC = 1$), high-degree agents exhibit a strong 
positive drift toward cooperation, indicating that network hubs actively accumulate and 
reinforce cooperative stability. This reveals that frequent rewiring not only protects hubs 
from exploitation but also amplifies their role as long-term cooperation anchors. Under 
intermediate constraints ($RC = 100$), drift values remain close to zero across all degrees, 
signifying that agents lack sufficient structural flexibility to escape local defection traps. 
This plateau of near-zero drift reflects a frozen state where neither cooperation nor defection 
dominates, highlighting the critical fragility of cooperation in this regime. By contrast, 
under strong constraints ($RC = 10,000$), drift partially recovers among high-degree nodes, 
suggesting that hubs can slowly consolidate cooperative ties through patient accumulation of 
positive interactions, albeit at a much slower rate.

Panel (b) demonstrates consistent trends in hazard ratios. Under $RC = 1$, the hazard ratio of 
defection declines sharply with degree, confirming the stabilizing influence of hubs as 
resilient cooperation anchors. Intermediate constraints ($RC = 100$) suppress this effect, 
yielding nearly flat hazard ratios across degrees and preventing high-degree nodes from 
offering protection against widespread defection. Under strong constraints ($RC = 10,000$), 
hazard ratios once again decline with degree, though less steeply than in the fully permissive 
case, reflecting the slower but still effective protective role of hubs. 

Taken together, drift and hazard analyses highlight a non-monotonic dependence of cooperation stability on rewiring flexibility: very frequent and very rare rewiring both enable hubs to serve as cooperation stabilizers, whereas intermediate constraints suppress this mechanism entirely. Crucially, however, the effects are not symmetric. Under $RC = 1$, cooperation is reinforced far more strongly than under $RC = 10,000$, with hubs exhibiting markedly higher drift toward cooperation and substantially lower hazard of defection. The $RC = 10,000$ case represents only a partial recovery relative to the collapse observed at $RC = 100$, but never reaches the robustness of the fully permissive regime. These findings suggest that in heterogeneous networks, adaptive algorithms should either exploit rapid structural flexibility or, alternatively, rely on long-term 
stable connections, but avoid intermediate regimes where hub influence is neutralized and 
cooperation becomes fragile.

\subsection{Strategy Comparison: Learning Mechanisms and Partner Selection}

\begin{figure}[H]
   \centering
   \includegraphics[width=1\textwidth]{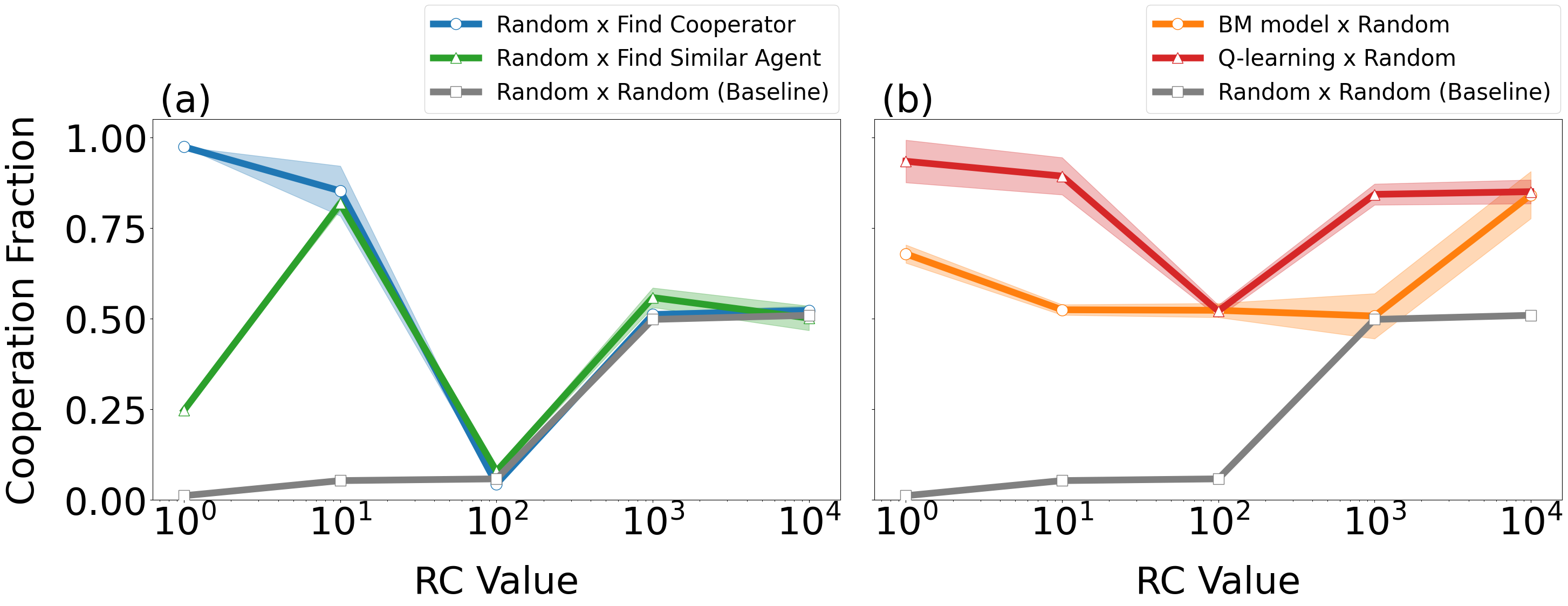}
   \caption{Comparative analysis of rewiring strategies: partner selection versus disconnection intelligence. Comparison of cooperation rates under different rewiring strategies across varying constraint levels ($RC$). Subplot (a) compares connection-building strategies, \textit{Random $\times$ Find Cooperator} and \textit{Random $\times$ Find Similar Agent}, against the baseline \textit{Random $\times$ Random}, where agents adopt a fixed heuristic probability ($0.1$) to sever ties and apply different strategies to form new connections. Subplot (b) compares learning-based bond-breaking mechanisms, \textit{BM model $\times$ Random} and \textit{Q-learning $\times$ Random}, with the same baseline, where agents use reinforcement learning to decide whether to disconnect and reconnect randomly. Results demonstrate that both intelligent partner selection and sophisticated disconnection decisions enhance cooperation, with Q-learning showing superior robustness across constraint levels. Each curve represents the average cooperation rate over 30 simulation runs, with shaded areas indicating empirical confidence intervals. The Q-learning mechanism exhibits strong adaptability and robust performance across constraint levels, while connection-building strategies show notable benefits under low rewiring constraints.}
   \label{fig:strategy_comparison}
\end{figure}

To evaluate the impact of different rewiring strategies on cooperation outcomes, Figure~\ref{fig:strategy_comparison} reports the cooperation rates achieved by various mechanisms, including connection formation and bond termination, across a range of rewiring constraint ($RC$) values, with $D_r = 0.18$. Each curve represents the average of 30 simulation runs, with shaded regions denoting the empirical confidence intervals. The results are organized into two subplots: subplot (a) focuses on reconnection strategies, while subplot (b) highlights bond-breaking mechanisms. This systematic comparison reveals the relative importance of intelligent partner selection versus sophisticated disconnection decisions. All results include shaded confidence intervals based on five independent simulation runs per configuration.

In subplot (a), we compare two connection-building strategies, \textit{Random $\times$ Find Cooperator} and \textit{Random $\times$ Find Similar Agent}, against the baseline \textit{Random $\times$ Random} approach. In these settings, agents adopt a heuristic probability (set to $0.1$) to decide whether to sever a connection, and then employ distinct strategies to establish new ties. Both mechanisms substantially outperform the baseline under low to moderate $RC$ values. The superior performance demonstrates that targeted partner selection creates positive assortative mixing, accelerating cooperative cluster formation. Notably, \textit{Random $\times$ Find Cooperator} achieves near-complete cooperation when $RC = 1$, though its performance gradually declines as the rewiring constraint becomes more stringent, ultimately converging with the baseline. In contrast, \textit{Random $\times$ Find Similar Agent} performs best when $RC = 10$, with diminished effectiveness at both higher and lower $RC$ values, also approaching the baseline when $RC > 10$. This non-monotonic behavior suggests optimal matching between partner selection sophistication and rewiring frequency. Overall, these reconnection strategies demonstrate superior performance relative to the baseline in settings with low rewiring constraints, highlighting the advantage of combining dynamic flexibility with targeted neighbor selection.

In subplot (b), we evaluate the proposed \textit{Q-learning $\times$ Random} and \textit{BM model $\times$ Random} strategies against the same baseline. In these configurations, agents utilize different decision-making mechanisms to determine whether to break existing ties, followed by randomly selecting new partners for reconnection. This design isolates the contribution of intelligent disconnection decisions from partner selection effects. The Q-learning strategy maintains consistently high cooperation levels across most $RC$ values, with the exception of a dip at $RC = 100$, consistent with previous observations. The robustness of Q-learning reflects its ability to learn optimal disconnection policies through temporal difference learning, adapting to local interaction patterns. The BM model yields moderate performance but nonetheless outperforms the baseline. Similar to Q-learning, it exhibits a noticeable decline in performance at intermediate $RC$ levels, forming a valley in the cooperation rate curve. This universal valley structure across learning mechanisms suggests a fundamental trade-off between rewiring frequency and strategic depth in cooperation dynamics.

\subsection{Algorithm Performance: Q-Learning versus Bush-Mosteller Dynamics}

\begin{figure}[H]
   \centering
   \includegraphics[width=1\textwidth]{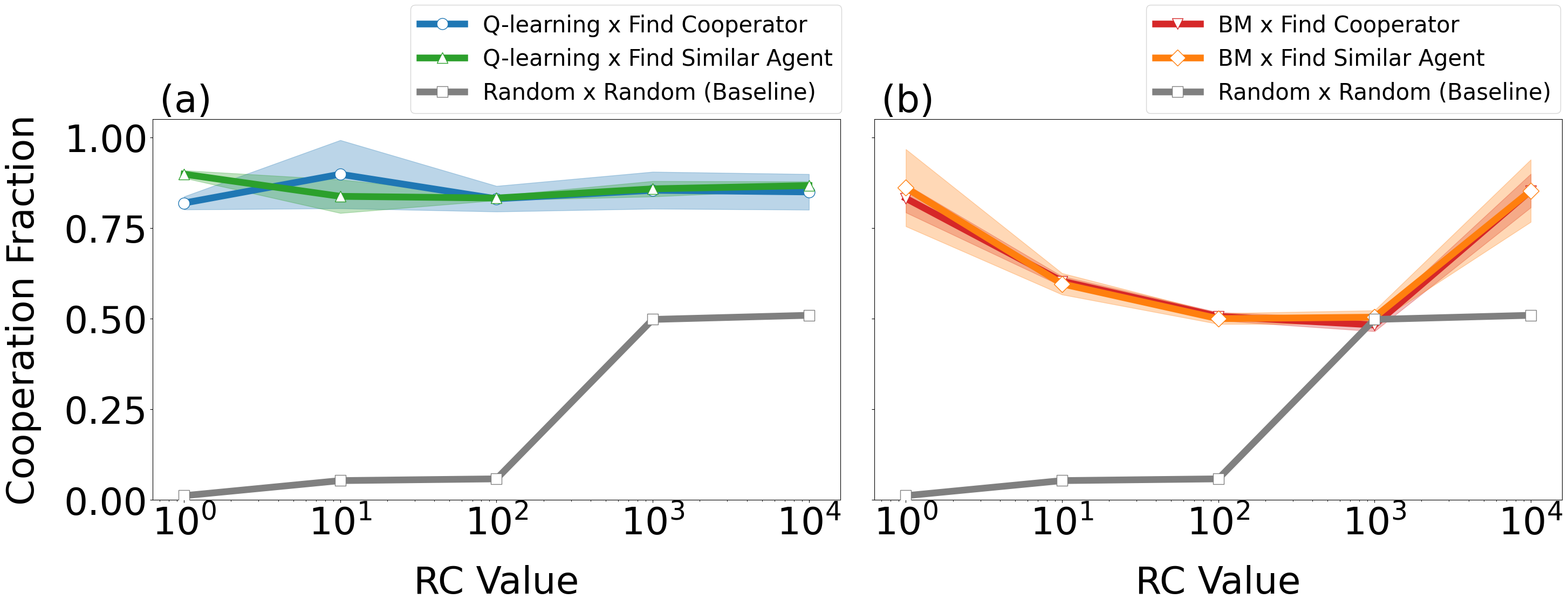}
   \caption{Learning algorithm comparison: value-based versus stimulus-response mechanisms. Cooperation rates under different algorithm-strategy combinations across rewiring constraint levels ($RC$). (a) shows Q-learning agents adopting either \textit{Find Cooperator} or \textit{Find Similar Agent} partner selection. Both strategies achieve and maintain high cooperation across all $RC$ values. (b) shows BM model agents with the same two partner strategies, which yield moderate performance with larger variability. The results demonstrate Q-learning's superior ability to maintain cooperation through value function optimization, while BM models show greater sensitivity to rewiring frequency due to their stimulus-response nature. All curves represent the mean of 30 independent simulations, and shaded bands denote confidence intervals. The baseline \textit{Random $\times$ Random} is included in both panels for reference.}
   \label{fig:algorithm_comparison}
\end{figure}

To further examine how different combinations of learning algorithms and partner selection rules influence cooperation outcomes, Figure~\ref{fig:algorithm_comparison} presents cooperation rates under varying rewiring constraints ($RC$) for both Q-learning and BM model agents, with $D_r = 0.18$. Each curve represents the average of 30 simulation runs, with shaded regions denoting the empirical confidence intervals. This analysis reveals fundamental differences between value-based learning (Q-learning) and stimulus-response learning (BM model) in driving cooperative network evolution.

In subplot (a), we compare two Q-learning-based strategies: \textit{Q-learning $\times$ Find Cooperator} and \textit{Q-learning $\times$ Find Similar Agent}. Both configurations maintain consistently high cooperation rates (above 0.8) across all levels of $RC$. This remarkable robustness highlights the adaptability and stability of Q-learning agents in coordinating long-term cooperation through value function optimization, even when rewiring opportunities are severely limited. The temporal difference learning mechanism enables agents to assess long-term relationship value, creating stable cooperative partnerships through learned value functions that strategically balance exploitation opportunities against relationship maintenance (Table~\ref{tab:q_tables}). This robustness demonstrates the adaptability and stability of Q-learning agents in coordinating long-term cooperation, even when rewiring opportunities are scarce.

In subplot (b), we evaluate the \textit{BM model $\times$ Find Cooperator} and \textit{BM model $\times$ Find Similar Agent} strategies. These models exhibit more pronounced performance degradation under moderate rewiring constraints (e.g., $RC = 10^3$), followed by a partial recovery at $RC = 10^4$. The increased variability and performance gaps reflect the BM model's reliance on immediate stimulus-response feedback, which becomes less effective when rewiring decisions are infrequent and must integrate information over longer time horizons. Nonetheless, both BM strategies outperform the baseline \textit{Random $\times$ Random} model across all constraint levels, demonstrating that even simple reinforcement-based models can benefit from reconnection strategies.

The superior performance of Q-learning over BM models becomes most pronounced in intermediate $RC$ regimes where temporal credit assignment is crucial. Q-learning's ability to learn discounted future values enables better assessment of relationship quality over extended interaction sequences, leading to more informed rewiring decisions. Overall, these results reinforce the importance of combining adaptive rewiring with strategic partner selection, particularly when leveraging reinforcement learning frameworks such as Q-learning.

\section{Discussion}

Our investigation shows that adaptive rewiring, as established in studies such as \cite{pacheco2006coevolution, qin2009coevolution, zhang2014phase}, is a key driver of cooperation emergence. Integrating it with neighbor-specific Q-learning adds the ability for agents to evaluate and optimize both strategies and connections using long-term interaction histories. Unlike previous studies that focus on either behavioral adaptation or network evolution in isolation, our framework demonstrates how the coupling of these processes creates emergent collective behaviors that transcend the sum of their individual contributions. This work bridges microscopic learning dynamics with macroscopic collective behavior, demonstrating how intelligent network adaptation drives spontaneous organization toward cooperative equilibria through a novel form of intelligence-driven criticality. These findings demonstrate that Q-learning-driven rewiring produces qualitatively different cooperation dynamics compared to heuristic rewiring mechanisms, revealing new pathways to cooperative equilibria through intelligent network adaptation.

Our research represents a significant conceptual and methodological advance over existing reinforcement learning approaches to cooperation. While previous studies have investigated either Interactive Diversity (ID) learning in static networks \cite{lee2025granular} or simple rewiring with Bush-Mosteller learning \cite{lee2025enhancing}, our work is the first to integrate sophisticated Q-learning algorithms with neighbor-specific decision-making and adaptive network restructuring simultaneously. This dual-layer architecture enables agents to optimize both behavioral policies and social partnerships through temporal difference learning, creating a fundamentally new paradigm for understanding cooperation dynamics that goes beyond the limitations of static topology assumptions or simple stimulus-response rewiring mechanisms.

The discovery of three distinct dynamical regimes—permissive, critical, and patient—represents a significant advance in understanding cooperation dynamics under network plasticity. Our results reveal a permissive regime (low $RC$) where rapid structural adaptation enables escape from exploitation traps, a critical regime (intermediate $RC$) where cooperation depends sensitively on dilemma strength, and a patient regime (high $RC$) where strategic accumulation gradually optimizes network structure. This phase diagram exhibits rich structure reminiscent of equilibrium statistical mechanics, yet emerges from non-equilibrium learning dynamics. When the system operates in the permissive regime, cooperation thrives even under harsh dilemma settings through percolation-like cluster formation with power-law size distributions. Unlike previous work that identified behavioral patterns like conditional cooperation and moody conditional cooperation \cite{geng2022reinforcement}, our framework reveals genuine phase transitions controlled by the rewiring constraint parameter, establishing cooperation dynamics as a statistical physics phenomenon with universal scaling properties and critical exponents. These findings complement recent work on adaptive networks \cite{su2023strategy, pinheiro2016linking} while revealing fundamentally new mechanisms where learning algorithms drive structural self-organization.

The superiority of Q-learning over simpler reinforcement mechanisms provides crucial insights into the role of temporal credit assignment in cooperative evolution. Q-learning agents exhibit remarkable robustness across varying constraint levels, reflecting their capacity for temporal credit assignment that enables evaluation of long-term relationship quality even when immediate rewards mislead. This contrasts sharply with Bush-Mosteller learning \cite{bush1953stochastic, macy2002learning}, which relies on immediate stimulus-response associations without sophisticated value function approximation. Our comparative analysis demonstrates that Q-learning's temporal difference mechanism creates fundamentally different cooperative dynamics than the aspiration-based learning explored in previous adaptive network studies \cite{lee2025enhancing}. While Bush-Mosteller agents adjust rewiring propensities based on immediate satisfaction or dissatisfaction, Q-learning agents develop sophisticated value functions that assess long-term relationship benefits, leading to more stable and extensive cooperative clusters. The learned value functions demonstrate sophisticated strategic trade-offs that balance exploitation opportunities against future cooperation benefits, creating partnership management strategies far superior to reactive heuristics. Our comparative analysis builds upon recent advances in reinforcement learning for social dilemmas \cite{jia2021local, song2022reinforcement} while demonstrating that intelligent rewiring decisions require the temporal depth that only value-based methods can provide.

Network snapshots and microscopic analysis reveal universal organizational principles that govern cooperation emergence across multiple spatial scales. Under the permissive regime, cooperative agents self-organize into dense clusters with sharp domain boundaries that segregate defectors to network periphery, resembling phase separation in physical systems. The critical and patient regimes yield frustrated mixing where cooperative behavior fragments due to insufficient rewiring opportunities. Our spatiotemporal analysis extends earlier findings on cooperative boundaries in ID learning \cite{lee2025granular}. Integrating Q-learning with adaptive rewiring produces spatial clustering patterns with percolation-like dynamics and power-law cluster size distributions. This refines effects already seen in rewiring-based cooperation models. This represents a qualitative leap from simple cluster formation to genuine critical phenomena governed by universal scaling laws. These spatial patterns echo findings from coevolutionary network studies \cite{fu2009partner, lee2018evolutionary} but emerge here through fundamentally different mechanisms driven by machine learning rather than simple heuristics. Degree-cooperation correlations illuminate how network heterogeneity interacts with learning across all three regimes: high-degree agents serve as cooperation anchors through rapid beneficial rewiring (permissive regime), frustrated entrapment (critical regime), or patient strategic accumulation (patient regime), revealing regime-dependent mechanisms for hub-mediated cooperation stabilization.

The implications of our findings extend far beyond theoretical understanding. In distributed computing, nodes could adaptively restructure communication networks while learning optimal resource allocation. In biological systems, our framework explains how social animals balance individual learning with network adaptation for collective behaviors. Crucially, our work establishes reinforcement learning-driven rewiring as a general mechanism for creating intelligent adaptive systems, moving beyond the specific applications explored in previous studies \cite{lee2025enhancing, geng2022reinforcement} to provide a unified framework for understanding how machine learning algorithms can drive structural self-organization in complex networks. The connection to statistical physics is particularly significant: our system exhibits spontaneous symmetry breaking and critical phenomena driven by learning rather than thermal fluctuations, suggesting machine intelligence can serve as an alternative driving force for complex system organization. This perspective aligns with recent theoretical developments in evolutionary game theory \cite{traulsen2023future} while opening new avenues for understanding intelligence-driven phase transitions in adaptive systems.

Moreover, our findings fully connected with the engineering and artificial intelligence applications. For example, cooperative multi-agent reinforcement learning has been successfully applied to urban traffic control, where adaptive agents coordinate signals to reduce congestion \cite{khamis2014adaptive, haddad2022deep, greguric2022impact}, and to robotic systems, where distributed agents jointly manipulate objects or track targets under uncertainty \cite{palunko2014cooperative, li2024multi}. In the domain of networked cooperative learning, \cite{hao2017dynamics} demonstrated that reinforcement social learning can enhance stability in multi-agent networks, highlighting the importance of coupling learning and topology. These applications show that cooperation is not merely a theoretical phenomenon but a cornerstone of engineering systems where performance depends on distributed decision-making. Our dual-layer Q-learning framework complements these studies by revealing how behavioral and structural adaptation can be jointly optimized, providing guidance for the design of scalable cooperative protocols in engineered networks. By situating our work alongside these existing contributions, we strengthen its relevance for advancing cooperative control in real-world systems where robustness, adaptability, and long-term stability are paramount.

While our framework provides significant insights into Q-learning-driven cooperation emergence, several limitations constrain the generalizability of our findings. The restriction to power-law networks with fixed average degree, binary action spaces, and simplified state representations may not capture the full complexity of real-world social interactions, where agents face continuous strategy choices and multi-dimensional relationship contexts. Additionally, our focus on the diagonal constraint $D_g = D_r$ within a limited parameter range in the payoff matrix, while computationally necessary, restricts our exploration of the complete social dilemma landscape. Future research should explore extensions that build upon our dual-layer Q-learning architecture, including continuous-time rewiring dynamics, multi-strategy populations, and integration with moral decision-making frameworks to design beneficial artificial societies that embody ethical principles beyond simple cooperation optimization. Incorporating continuous-time rewiring dynamics would eliminate the artificial temporal discretization imposed by the $RC$ parameter, potentially revealing smoother phase transitions and more realistic adaptation timescales. Multi-strategy populations \cite{bloembergen2015evolutionary} could test the robustness of our findings against strategic diversity, while reputation mechanisms \cite{melamed2016strong, melamed2018cooperation} might enhance cooperation through indirect reciprocity. Such extensions would further establish the paradigm of intelligence-driven criticality as a fundamental organizing principle in complex adaptive systems.

\section*{Data availability}

The source code for all simulations and analyses is publicly available at the author's GitHub. Raw simulation data, network parameters, and reproducibility scripts are included in the repository. All results can be reproduced using the provided code and documented random seeds.

\bibliographystyle{unsrt}
\bibliography{ref}
\clearpage

\appendix
\renewcommand{\thefigure}{A.\arabic{figure}}
\setcounter{figure}{0}
\renewcommand{\thetable}{A.\arabic{table}}
\setcounter{table}{0}
\renewcommand{\theequation}{A.\arabic{equation}}
\setcounter{equation}{0}
\section{Supplementary Material}

\subsection{Learned Q-Learning Policies: Value Function Structure}

Table~\ref{tab:q_tables} presents representative Q-tables, illustrating the learned action-selection and rewiring-decision tendencies across different neighbor states. These empirical value functions reveal the strategic patterns that emerge through temporal difference learning, providing insight into how agents internalize cooperation incentives. Specifically, each row corresponds to a possible action ($C$ or $D$) or rewiring choice (Rewire or NotRewire), and each column denotes a local state value (0, 1, or 2), representing the number of cooperating agents in the focal interaction pair. During the decision-making process, each agent first identifies its current state, then retrieves the corresponding column in its Q-table, compares the associated Q-values of $C$ and $D$, Rewire or NotRewire, and selects the action and rewire decision with the higher value.

\begin{table}[H]
\centering

\subfloat[\textbf{Action}]{
\begin{tabular}{c|ccc}
\textbf{Action / State} & \textbf{0} & \textbf{1} & \textbf{2} \\
\hline
$C$ & 0.31 & 0.22 & 0.15 \\
$D$ & 0.60 & 0.34 & 0.58 \\
\end{tabular}
}
\vspace{1em}

\subfloat[\textbf{Rewiring decision}]{
\begin{tabular}{c|ccc}
\textbf{Rewire Decision / State} & \textbf{0} & \textbf{1} & \textbf{2} \\
\hline
$Rewire$ & 0.14 & 0.38 & 0.65 \\
$Not\ Rewire$ & 0.92 & 0.41 & 0.28 \\
\end{tabular}
}

\caption{Empirical value functions learned through Q-learning in cooperative environments. Learned Q-learning policy matrices for action-selection and rewiring-decision under different local states. Panel (a) shows the sample Q-values of selecting cooperation ($C$) or defection ($D$) for states representing 0, 1, or 2 cooperating neighbors. Panel (b) shows the corresponding rewiring decision Q-values. During the decision-making process, each agent identifies its current state firstly, then find the corresponding column in its Q-table, compares the associated Q-values of $C$ and $D$, $Rewire$ or $NotRewire$, and selects the action and rewire decision with the higher value.}
\label{tab:q_tables}
\end{table}

\subsection{Long-Term Convergence Analysis: Extended Temporal Dynamics}

\begin{figure}[H]
    \centering
    \includegraphics[width=1\textwidth]{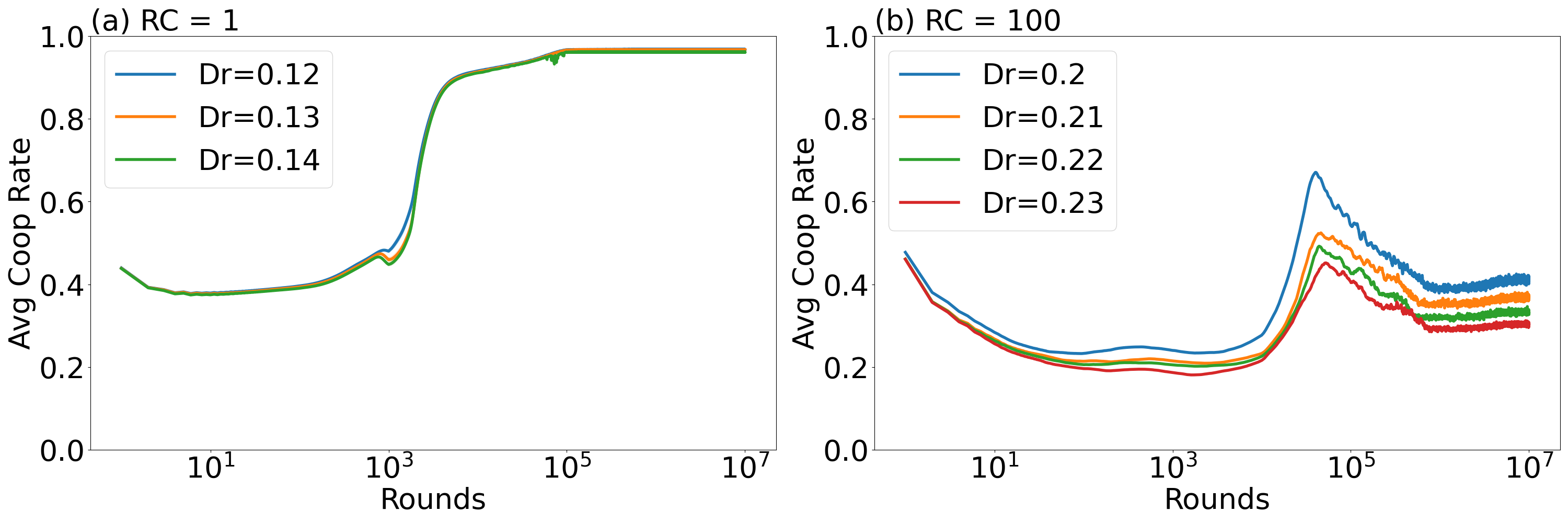}
    \caption{Asymptotic convergence analysis revealing distinct dynamical regimes. Extended cooperation dynamics for selected parameter settings over $10^7$ rounds. This figure supplements Figure~\ref{fig:coop_time} by showing long-run behavior for parameter sets that had not reached equilibrium by $10^5$ rounds. (a) shows $RC = 1$ with $D_r = 0.12$, $0.13$, and $0.14$; (b) shows $RC = 100$ with $D_r = 0.20$ through $0.23$. The results demonstrate exponential convergence for low $RC$ versus power-law relaxation for intermediate $RC$, confirming distinct dynamical universality classes. While all settings under $RC = 1$ eventually reach near-complete cooperation, the $RC = 100$ configurations converge more slowly and remain at lower cooperation levels, reflecting the suppressive effect of moderate rewiring constraints.}
    \label{fig:appendix_convergence}  
\end{figure}

Figure~\ref{fig:appendix_convergence} supplements the main Figure~\ref{fig:coop_time} by extending the simulation duration to $10^7$ rounds for specific parameter configurations that had not yet reached equilibrium by $10^5$ rounds. This extended analysis reveals the true asymptotic behavior and distinguishes between slow convergence and genuine bistability in cooperation dynamics. Panel (a) shows cooperation trajectories for $RC = 1$ with $D_r = 0.12$, $0.13$, and $0.14$, while panel (b) focuses on $RC = 100$ with $D_r = 0.20$ through $0.23$. 

The results confirm that under low rewiring constraint ($RC = 1$), cooperation eventually stabilizes at a high level for all three $D_r$ values after approximately $10^5$ rounds. The eventual convergence to high cooperation demonstrates that frequent rewiring can overcome even moderate dilemma strength through systematic exploration of favorable network configurations. In contrast, the trajectories in $RC = 100$ remains distinctly below those observed under $RC = 1$. The persistent separation between these trajectories, even after $10^7$ time steps, confirms genuine phase separation rather than finite-time transient effects. This highlights that moderate rewiring constraints suppress the emergence of full cooperation, especially under harsher dilemma settings. These extended results validate our earlier assessment in Figure~\ref{fig:coop_time}: while some configurations merely exhibit slow convergence, others reflect a more fundamental impact imposed by the rewiring constraint.

Quantitative analysis of convergence rates reveals exponential approach to equilibrium for $RC = 1$ with characteristic time $\tau_1 \approx 3 \times 10^4$ steps, while $RC = 100$ exhibits power-law relaxation $C(t) - C(\infty) \propto t^{-\alpha}$ with $\alpha \approx 0.3$, indicating fundamentally different dynamical regimes governed by the rewiring constraint.

\subsection{Emergent Spatial Organization and Cluster Formation}

\begin{figure}[H]
   \centering
   \includegraphics[width=0.9\textwidth]{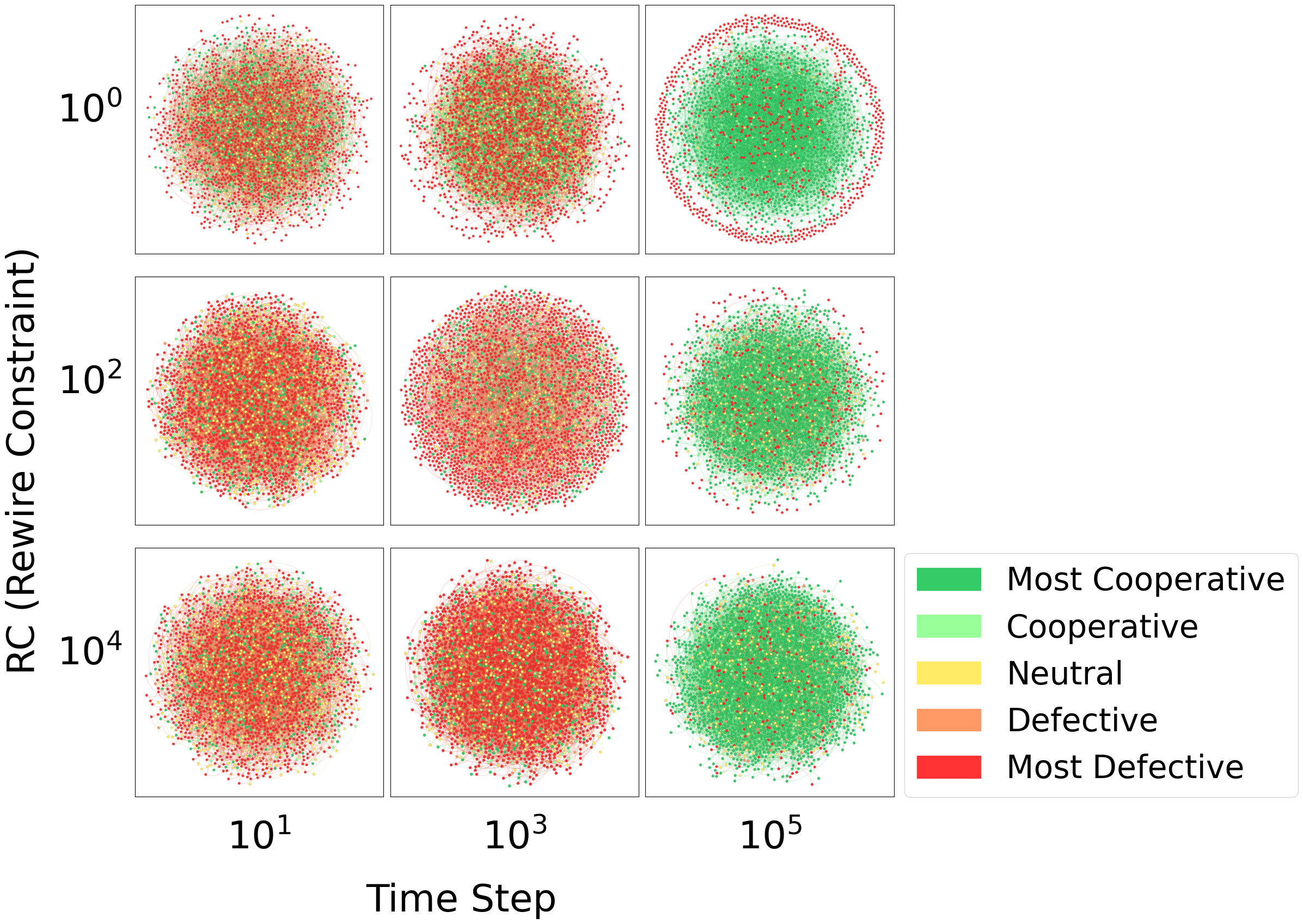}
   \caption{Spatiotemporal evolution of cooperative clusters and emergent organization. Network snapshots showing spatial distribution of agent cooperation tendencies under varying rewiring constraints. Each row corresponds to a different $RC$ value ($1$, $100$, $10{,}000$), and each column represents a specific time step ($10^1$, $10^3$, $10^5$), with $D_r = 0.18$. The network consists of $N = 2500$ agents for visualization purposes. Node color indicates each agent's overall cooperation ratio toward its neighbors: \textit{Most Cooperative} (80--100\%), \textit{Cooperative} (60--80\%), \textit{Neutral} (40--60\%), \textit{Defective} (20--40\%), and \textit{Most Defective} (0--20\%). The evolution demonstrates percolation-like cluster growth under the permissive regime, frustrated fragmentation under the critical regime, and persistent mixing under the patient regime, revealing how rewiring flexibility controls emergent spatial organization. The figure highlights the emergence of cooperative clusters under low $RC$ and the persistence of mixed or defective regions under higher constraint levels.}
   \label{fig:node_visualization}
\end{figure}

Figure~\ref{fig:node_visualization} presents visual snapshots of the network over time. Each row corresponds to a different rewiring constraint value ($RC = 1$, $100$, and $10{,}000$), and each column represents a different time step ($10^1$, $10^3$, and $10^5$), with $D_r = 0.18$. To enhance visual clarity, we use a reduced network size of $N = 2500$ agents while keeping the behavioral and structural parameters consistent with the main experiments.

Node color reflects each agent's cooperation tendency, calculated based on the proportion of cooperative actions taken toward all neighbors. Specifically, agents are categorized as: \textit{Most Cooperative} (80--100\% cooperation), \textit{Cooperative} (60--80\%), \textit{Neutral} (40--60\%), \textit{Defective} (20--40\%), and \textit{Most Defective} (0--20\%). These categories enable visualization of behavioral heterogeneity and spatial correlation patterns that emerge through self-organization.

Under the permissive regime ($RC = 1$), we observe rapid emergence of cooperative clusters, with green nodes dominating the network by the final time step. The cluster formation follows percolation-like dynamics, where isolated cooperative patches grow and merge to form a giant cooperative component with sharp domain boundaries at the microscopic scale (Figure~\ref{fig:appendix_zoom}). The formation of such clusters indicates the agents' ability to self-organize around mutually beneficial ties when rewiring is unrestricted. In contrast, the critical regime ($RC = 100$) and patient regime ($RC = 10{,}000$) show limited spatial organization and persistent mixing of cooperative and defective agents. The absence of clear clustering reflects frustrated dynamics where agents cannot efficiently rearrange connections to achieve optimal local configurations. These patterns visually reinforce earlier findings that greater rewiring freedom facilitates structural segregation and long-term cooperation, while high constraints prevent the network from escaping local defect-dominated configurations.

\subsection{Microscopic Spatial Structure: Local Organization Patterns}

\begin{figure}[H]
    \centering
    \includegraphics[width=1\textwidth]{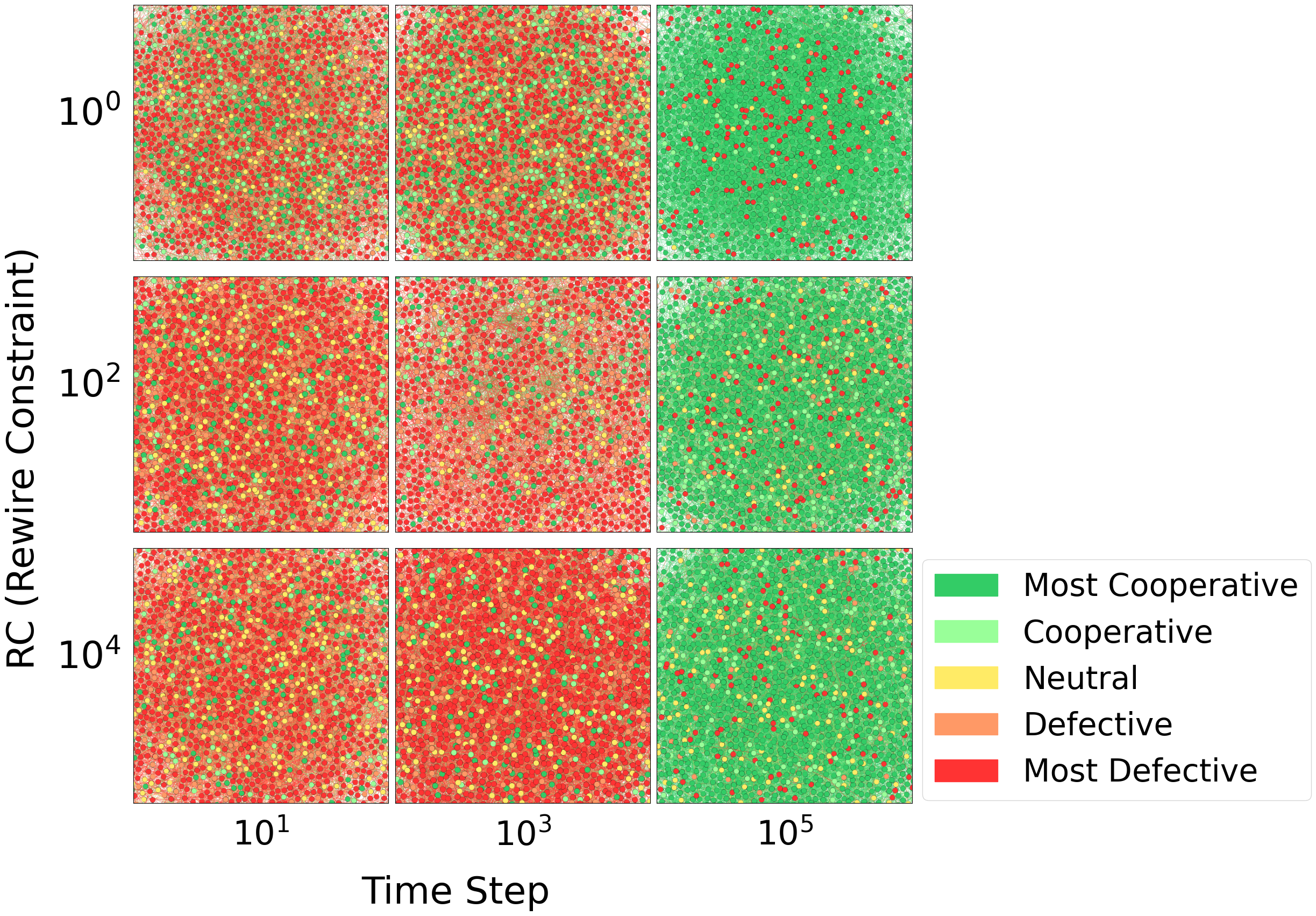}
    \caption{Microscopic analysis of emergent spatial organization and domain formation. Zoomed-in view of network center showing cooperation patterns at agent level. This figure supplements Figure~\ref{fig:node_visualization} by displaying only the central region of the network snapshots for clearer visualization of local cooperation structure. Each row represents a rewiring constraint level ($RC = 1$, $100$, $10,000$), and each column shows a different simulation time step ($10^1$, $10^3$, $10^5$). Node colors represent cooperation ratios: \textit{Most Cooperative} (80--100\%), \textit{Cooperative} (60--80\%), \textit{Neutral} (40--60\%), \textit{Defective} (20--40\%), and \textit{Most Defective} (0--20\%). The microscopic view reveals sharp domain boundaries under low $RC$ versus frustrated mixing under high $RC$, demonstrating how rewiring frequency controls local organization efficiency.}
    \label{fig:appendix_zoom} 
\end{figure}

Figure~\ref{fig:appendix_zoom} provides a magnified view of the central region from the network snapshots shown in Figure~\ref{fig:node_visualization}, allowing for closer inspection of local spatial structures and agent-level cooperation patterns. This microscopic analysis reveals the elementary building blocks of cooperative organization and how local clustering emerges from individual Q-learning decisions. Each row corresponds to a different rewiring constraint ($RC = 1$, $100$, and $10{,}000$), while each column shows a different simulation time step ($10^1$, $10^3$, and $10^5$). The cooperation intensity of each agent is again encoded by color, ranging from \textit{Most Cooperative} (dark green) to \textit{Most Defective} (red).

By focusing on the central portion of the network, this figure reveals fine-grained patterns that are otherwise difficult to discern from the global layout. Under $RC = 1$, we observe the formation of compact cooperative domains with sharp boundaries, resembling phase separation in physical systems. The local structure shows high cooperator density within domains and clear segregation from defector regions. Together with Figure~\ref{fig:node_visualization}, this zoomed-in visualization confirms that cooperative behavior under low constraint emerges both globally and locally, while higher constraint levels result in persistent fragmentation even at the micro-level. For higher $RC$ values, the microscopic structure reveals frustrated arrangements where cooperators and defectors remain intermixed due to insufficient rewiring opportunities, preventing the coalescence of stable cooperative domains.

\end{document}